# Intrinsic Josephson junctions in the iron-based multi-band superconductor $(V_2Sr_4O_6)Fe_2As_2$

Philip J.W. Moll[1*], Xiyu Zhu[2], Peng Cheng[2+], Hai-Hu Wen[2], Bertram Batlogg[1]

[1]*Physics of New Materials, Solid State Physics, ETH Zurich, CH-8093 Zurich, Switzerland*
[2]*Center for Superconducting Physics and Materials, National Laboratory of Solid State Microstructures and Department of Physics, Nanjing University, Nanjing 210093, China*
[+]*Present address: Physics Department, Renmin University, China*

*correspondence address: phmoll@phys.ethz.ch

**In layered superconductors, Josephson junctions may be formed within the unit cell[1–3] due to sufficiently low interlayer coupling. These intrinsic Josephson junction (iJJ) systems[4] have attracted considerable interest for their application potential in quantum computing as well as efficient sources of THz radiation, closing the famous "THz gap"[5]. So far, iJJ have been demonstrated in single-band, copper-based high-$T_c$ superconductors, mainly in Ba-Sr-Ca-Cu-O[6–10]. Here we report clear experimental evidence for iJJ behavior in the iron-based superconductor $(V_2Sr_4O_6)Fe_2As_2$. The intrinsic junctions are identified by periodic oscillations of the flux flow voltage upon increasing a well aligned in-plane magnetic field[11]. The periodicity is well explained by commensurability effects between the Josephson vortex lattice and the crystal structure, which is a hallmark signature of Josephson vortices confined into iJJ stacks[12,13]. This finding adds $(V_2Sr_4O_6)Fe_2As_2$ as the first iron-based, multi-band superconductor to the copper-based iJJ materials of interest for Josephson junction applications, and in particular novel devices based on multi-band Josephson coupling may be realized.**

Applications based on Josephson junctions, such as SQUID magnetometers, exploit the phase coherence of the macroscopic wavefunction of a superconductor. In these successful applications, artificially grown heterostructures incorporating a layer of order parameter suppression – typically an oxide barrier – serve as the Josephson junctions. With the discovery of cuprate high-$T_c$ superconductors and their small out-of-plane coherence length resulting from insulating layers separating the Cu-O planes, the idea of intrinsic Josephson junctions inside the crystal structure and consequently devices based on self-assembled nano-junctions quickly emerged.

A particular application example is the efficient generation of Terahertz radiation due to the ac-Josephson effect. In between the realms of solid state emitters at lower frequencies and optical emitters at higher frequencies, a frequency range around a few THz lacks efficient, compact and cost-effective emitters and detectors. Closing this blank spot for sensing and imaging applications, often referred to as the "THz gap", would be of significant practical interest[5] in biology, chemistry, material science as well as in security and safety applications.

Notable progress has been made in recent years towards the implementation of iJJs as emitters and detectors in the low THz range. Up to now all such emitters are based on cuprates, mainly on the highly anisotropic Ba-Sr-Ca-Co-O (BSCCO). With the advent of iron-based superconductors, the iron-pnictides, as a different class of layered high-$T_c$ superconductors, the question of their potential as iJJ systems arises naturally. While most structure classes of iron-pnictides are nearly isotropic with a strong interlayer coupling, the recent discovery of a transition to Josephson-like vortices in SmFeAs(O,F)[14] showed the possibility to observe Josephson vortices in the pnictides. SmFeAs(O,F), however, is only weakly anisotropic and its c-axis resistivity is too low to sustain a voltage difference between adjacent layers without significant interlayer currents. This started the search for even more strongly layered systems with weaker interlayer coupling leading to true iJJ behavior.

One obvious candidate of the pnictide families is $(V_2Sr_4O_6)Fe_2As_2$, a member of the "42622" structural class, due to its large layer spacing of $d_c$ = 1.56nm[15]. Of particular interest for application is the physical realization of an intrinsic Josephson system in a multiband superconductor. The multi-band nature of the iron-based superconductors is expected to influence the Josephson junction behavior, thus novel effects and applications have been proposed that are inaccessible for single-band superconductors such as cuprates: Josephson vortices in intrinsic $s_\pm$ junctions are predicted to show a significantly extended core region compared to single-gap materials due to destructive interference of the cooper pair tunneling channels[16]. Therefore the vortex dynamics are distinct from single-band Josephson systems due to the different pinning of multi-band Josephson vortices. Additionally, the interactions of the different tunneling channels will influence the interlayer coupling, potentially assisting the synchronization of junctions in a stack[17]. The differences in phase evolution around the phase-core region of a Josephson vortex, that arise from the additional inter- and intra-band tunneling channels, may also be exploited in phase sensitive iJJ applications such as quantum computing[9]

Here we show clear evidence for intrinsic Josephson behavior in the multi-band system $(V_2Sr_4O_6)Fe_2As_2$. The influence of the material's layeredness on the vortex matter and on its behavior as an intrinsic Josephson system can be best studied by electric transport measurements perpendicular to the layers, i.e. along the c-direction. Due to the small size and unfavorable plate-like shape of available single crystals, the samples were microfabricated and contacted using the FIB technique[18]: In this process, an individual crystallite was selected and carved into its final form shown in Figure 1a) by a $Ga^{2+}$ ion beam: The crystal (purple) has been carved into a 4-point geometry, and was contacted by FIB-platinum deposition (blue). This technique allows reliable measurements of the out-of-plane resistivity $\rho_c$ that is otherwise inaccessible to transport experiments in micro-crystallites.

Figure 1c) shows the temperature dependence of the c-direction resistivity $\rho_c(T)$. The resistivity increases from 29 mΩ cm at room temperature to 131 mΩ cm at $T_c \sim 19.8K$ following a power law dependence $\rho_c(T) \sim T^{-0.5}$. Such a temperature dependence of the out-of-plane resistivity is experimental evidence for the incoherent electron transport through insulating $Sr_2VO_3$ layers. These layers have been predicted to be insulating by LDA+U calculations[19] due to V-d orbital correlations leading to a Mott-like insulator, as confirmed by photoemission[20]. Therefore in this material SC-insulator-SC (SIS) junctions are expected instead of SC-Normal metal-SC (SNS)-type junctions, in agreement with our out-of-plane resistivity measurements.

This overall temperature dependence of $\rho_c$ is qualitatively reminiscent of the more anisotropic cuprates, however the absolute value of $\rho_c(T_c)$ is two orders of magnitude smaller than the highly anisotropic iJJ compounds $Bi_2Sr_2CaCu_2O_{8+x}$ (10-30 Ω cm) and $Tl_2Ba_2Ca_2Cu_3O_{10}$ (13 Ω cm)[3], but higher than the more isotropic pnictide SmFeAs(O,F) (2.7 mΩcm)[18]. This comparatively lower out-of-plane resistivity in $(Sr_4V_2O_6)Fe_2As_2$ is indicative of a stronger interlayer coupling than in $Bi_2Sr_2CaCu_2O_{8+x}$.

The key observation clearly revealing the intrinsic Josephson effect is the vortex matter behavior when the magnetic field is well-aligned parallel to the planes and a dc-current flows perpendicular to the layers (Figure 2). In this configuration, the Lorentz force induces an in-plane sliding motion of vortices parallel to the superconducting layers as sketched in Figure 2a). Measurable flux flow sets on just above 1T and increases monotonically with increasing field. In fields above 2.5T, well defined periodic oscillations appear ontop of an increasing background that persist to fields above 6T. The oscillation period is constant over the whole field range with a periodicity of $H_P \approx 0.178T$, and the oscillations persist over an extended temperature window from 15K down to 2K, the lowest accessible temperature in the experiments.

This periodic modulation of the out-of-plane flux-flow voltage upon increasing a well-aligned in plane field is a hallmark of intrinsic Josephson behavior. Such oscillatory effects have been extensively studied experimentally in cuprate intrinsic Josephson compounds[6,11,21,22] and their origin has been well identified theoretically as a result of a periodic modulation of the surface barrier due to vortex matching effects[23,24]. For magnetic fields well aligned with the superconducting layers, vortices in an iJJ system are purely of Josephson character without any Abrikosov-like segments ("pancake vortices") in the superconducting layers. In this case, vortex motion perpendicular to the layers, i.e. across the superconducting barrier, is strongly suppressed and this intrinsic vortex barrier is sometimes referred to as intrinsic pinning[2]. The Josephson vortices, however, are much weaker pinned by defects due to the absence of a normal core[25], and thus are highly mobile between two adjacent FeAs planes[14]. Without effective defect pinning, the main source of pinning is the surface barrier that impedes vortex entry and exit. This surface barrier is known to be modulated by commensurability effects between the layered crystal structure and the Josephson vortex lattice, and thus gives rise to the observed oscillatory phenomenon.

The periodicity $H_P$ corresponds to the field required to add one flux quantum per two unit cell layers, due to a hexagonal arrangement of the vortices. The vortex structure associated with these commensurability oscillations is sketched in Figure 2b) for the two exemplary cases of 1 and 1.5 flux quanta per layer. The two configurations differ by one additional vortex in every second layer, and the transition between them requires a change by one flux quantum per area $(2d_c) * w$, where w denotes the sample width. Therefore the oscillation period depends on the material specific interlayer distance $d_c$ as well as on the sample dependent geometry, as has been clearly shown in cuprates[11]:

$$H_P = \phi_0 / 2wd_c \quad (1)$$

Using the c-axis lattice parameter determined by X-ray diffraction as $d_c$ = 1.567 nm[15] and the width of the FIB-cut sample measured by SEM as w = 3.6µm, the calculated periodicity field $H_P = 0.177 T$ is in perfect agreement with the observed periodicity of 0.178T. The oscillations are expected to occur only at high enough vortex densities, in which the Josephson junctions are homogeneously penetrated by vortices. This penetration field has been theoretically predicted[23] as $H_{cr} = \Phi_0 / 2\pi\gamma d_c^2$, where γ denotes the coherence length anisotropy. As the onset of oscillations is observed at about 2.5T, the anisotropy can thus be estimated as $\gamma \approx 51$. This value is significantly lower than typical values of $\gamma > 1000$ in $Bi_2Sr_2CaCu_2O_{8+x}$, in line with the significantly lower out-of-plane resistance in $(V_2Sr_4O_6)Fe_2As_2$.[3]

An important aspect of the Josephson oscillations is their robustness against materials defects, clearly excluding extrinsic Josephson junctions, like planar intergrowth layers, as the origin of these observations. In all studied samples, a small non-zero resistance was observed below $T_c$, most likely as a result of layered defects. Such layered defects are common in long c-axis materials, and highlight the complex chemistry and synthesis challenges in this quinary oxide. Due to this finite resistance in the SC state, self-heating prevented the observation of branches in the I-V characteristics, another key evidence for intrinsic Josephson junctions[26–28]. These defects, however, do not contribute to the oscillatory signal unless they occur in periodic arrays along the c-direction and are all of the exact same thickness. The absence of additional frequencies, as shown in the inset of Figure 2, and the clear coincidence of the observed effective Josephson junction thickness with the material parameter $d_c$ clearly exclude this highly unlikely scenario.

The similarity of the oscillatory enhancement of flux flow dissipation between $(V_2Sr_4O_6)Fe_2As_2$ and $Bi_2Sr_2CaCu_2O_{8+x}$ is evident from Figure 3. The experimentally observed oscillation periods in both compound families well follow the linear relation given by eq.(1). This indicates a similar Josephson vortex arrangement among these very different classes of superconductors. Interestingly, the frequencies in BSCCO are observed slightly below the theoretically predicted $H_P$, while the periodicity in $(V_2Sr_4O_6)Fe_2As_2$ matches the predicted periodicity very well. As $(V_2Sr_4O_6)Fe_2As_2$ is less anisotropic, this may indicate a more coherent Josephson behavior along the stack.

In conclusion, we have shown intrinsic Josephson junctions in the iron-based superconductor $(V_2Sr_4O_6)Fe_2As_2$, and thereby added the first multi-band material to the group of compounds suitable for iJJ applications such as THz emitters and qubit implementations. In addition, its multi-band nature gives rise to new quantum coherent processes in the junction due to inter- and intra-band tunneling channels. This additional degree of freedom in the tunneling process maintaining the phase coherence through the stack is expected to have profound impact on the junction characteristics, such as an elongation of the core region of Josephson vortices compared to single-gap superconductors[16]. Thus the intrinsic Josephson behavior in $(V_2Sr_4O_6)Fe_2As_2$ may lead to new types of iJJ applications that cannot be realized in cuprate superconductors.

**Methods**

**Crystal growth** - The high-quality single crystals of $Sr_4V_2O_6Fe_2As_2$ were grown by the flux method. First the starting materials $V_2O_5$ powder (purity 99.95%), $Fe_2O_3$ powder(purity 99.9%), Sr grains (purity 99%) and FeAs powder were mixed in stoichiometry according to $Sr_4V_2O_6Fe_2As_2$. Then the powder was ground with more FeAs powder of molar ratio (FeAs : $Sr_4V_2O_6Fe_2As_2$ = 2: 1), sealed in an evacuated quartz tube and heated at about 1150 °C for 2 days. Then it was cooled at a rate of 3 °C/hour to 1050 °C and followed by a quick cooling down by shutting off the power of the furnace. The resulting product is a pellet with dark color that contains mainly the plate-like small crystals.

**Measurements** – Resistivity measurements were performed in quasi-dc mode by applying rectangular current pulses. For all measurements, a Keithley Nanovolt option was used in the „Delta Mode" setting, thus switching the current from +I to –I every 83ms. The field was well aligned with the FeAs-layers using an Attocube slip-stick rotator by maximizing the flux-flow motion in constant applied field.

# Figures

## Figure 1 : The intrinsic Josephson pnictide (Sr$_4$V$_2$O$_6$)Fe$_2$As$_2$

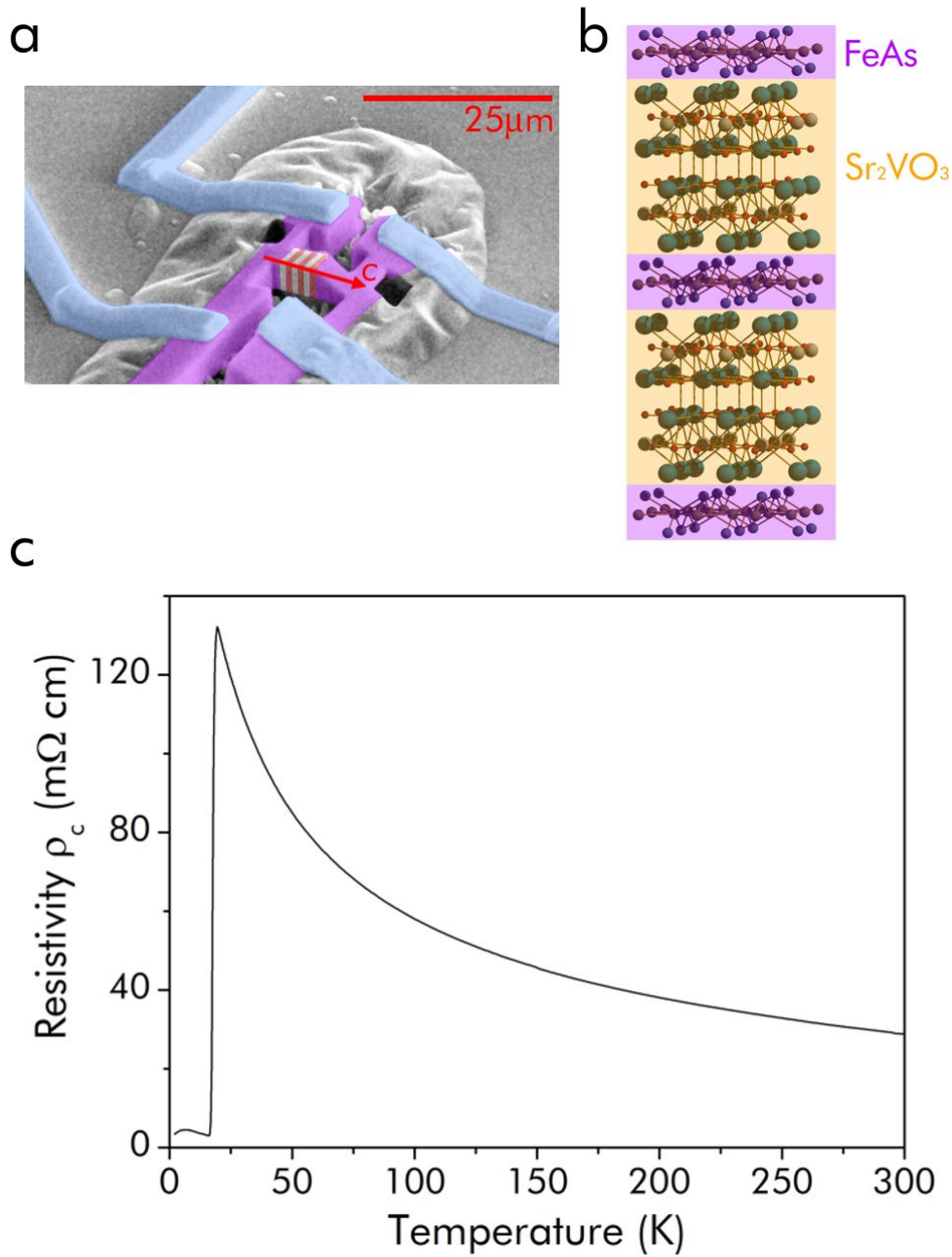

a) SEM micrograph of the microstructured crystal of (V$_2$Sr$_4$O$_6$)Fe$_2$As$_2$ (purple). The c-direction resistivity is measured in a four point geometry, that was carved out of a singly crystal by Focused Ion Beam (FIB) cutting and contacted via FIB-deposited Pt-leads (blue). The active Josephson junction stack is in between the voltage contacts, and the direction of the layers is indicated (red&white planes).
b) Crystal structure of (V$_2$Sr$_4$O$_6$)Fe$_2$As$_2$, depicting the intrinsic tunnel barrier VSr$_2$O$_3$-blocks in between the superconducting FeAs layers.
c) C-direction resistivity of (V$_2$Sr$_4$O$_6$)Fe$_2$As$_2$ as a function of temperature. Upon cooling, the temperature increases indicating the insulating nature of the VSr$_2$O$_3$-blocks. The transition temperature $T_c$ is 19K, indicating slightly underdoped samples $(V_2Sr_4O_{2.2})Fe_2As_2$[29].

## Figure 2: Josephson vortex commensurability oscillations

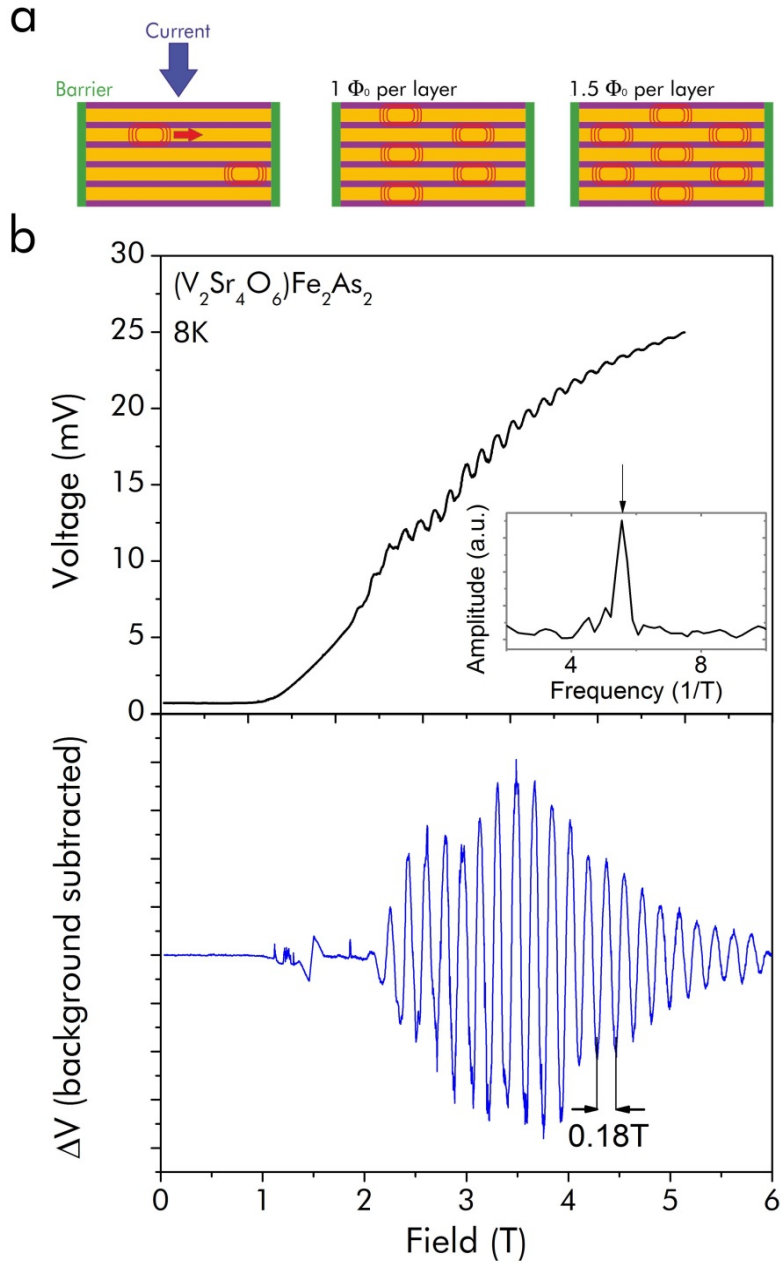

a) Sketch of the vortex motion in the iJJ stack. The out-of-plane current drives the Josephson vortices in a sliding motion through the structure. While point pinning is ineffective for Josephson vortices, the surface barrier impedes vortex entry or exit. The middle and right panels show a sketch of two neighboring commensurate vortex configurations. Upon increasing the field from the middle configuration, $1\ ^{\Phi_0}/_{layer}$, the next commensurate configuration is reached at $1.5\ ^{\Phi_0}/_{layer}$. Therefore a periodicity of one flux quantum per two unit layers corresponds to the observed vortex lattice periodicity.

b) Flux-Flow voltage at constant dc-current of 20µA as a function of in-plane magnetic field. At high fields above 2.5T, pronounced oscillations appear on the otherwise monotonically increasing background of flux motion. These oscillations are periodic in field, and only one significant periodicity is observed (Inset: FFT of the signal).

## Figure 3 : Comparison to copper-based intrinsic Josephson junction systems

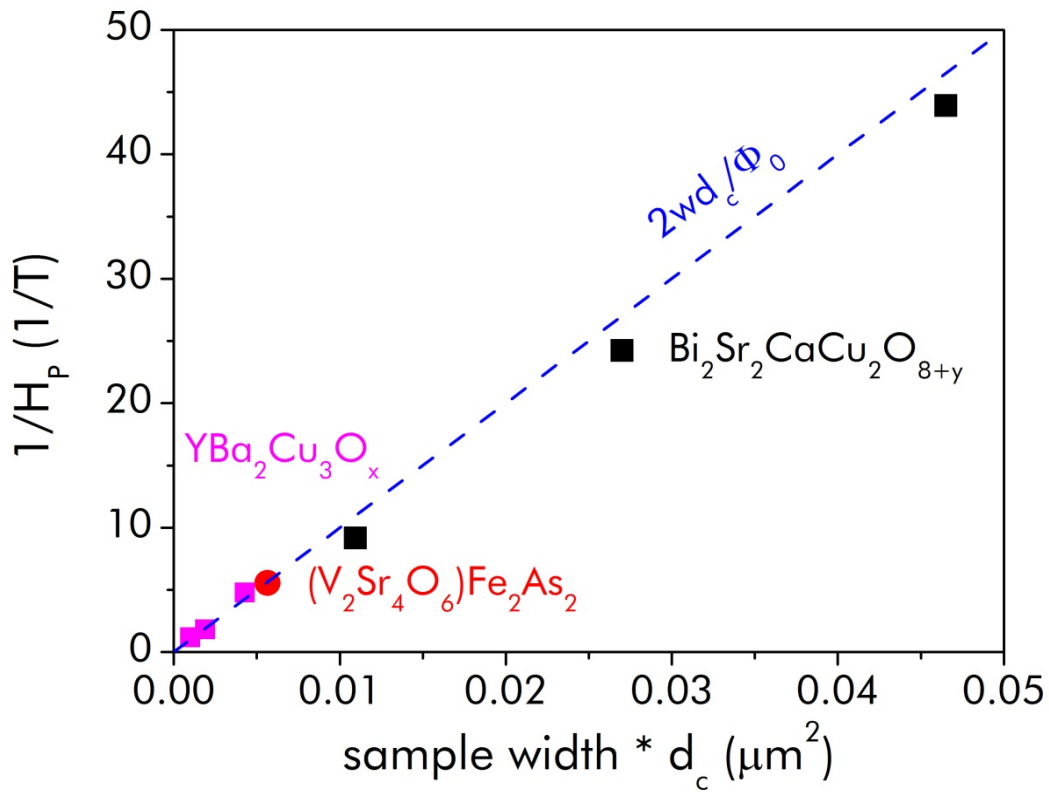

Comparison of the field periodicity of the intrinsic Josephson oscillations from the iron-pnictide $(V_2Sr_4O_6)Fe_2As_2$ (this work, red) with the cuprate systems $Bi_2Sr_2CaCu_2O_{8+y}$ (black, from ref[11]) and $YBa_2Cu_3O_x$ (magenta, from ref[22]). All of these different compounds show Josephson oscillation periods in good agreement with the theoretically expected $1/H_P = 2wd_c/\Phi_0$. This provides evidence that similar interlayer Josephson vortex arrangements are realized in $(V_2Sr_4O_6)Fe_2As_2$ and Cuprate superconductors, despite their distinct physics.